 \documentclass[preprint,5p]{elsarticle}
 \usepackage[utf8]{inputenc}
 \usepackage[T1]{fontenc}

 \usepackage{graphics}

\usepackage{amssymb}
\usepackage{makeidx}
\usepackage{graphicx}
\usepackage{color}
\usepackage{multicol}
\usepackage[bottom]{footmisc}
\usepackage[english]{babel}
\usepackage{amsmath}
\usepackage{xcolor}
\definecolor{link_blue}{RGB}{52,46,157}
\usepackage[colorlinks,citecolor=link_blue,linkcolor=link_blue,urlcolor=link_blue]{hyperref}

\biboptions{sort&compress}
\DeclareMathOperator{\re}{Re}
\DeclareMathOperator{\im}{Im}

\journal{NIM B}

\begin{document}

\begin{frontmatter}



\title{Radical increase of the parametric X-ray intensity under condition of extremely asymmetric diffraction}


\author[label6]{O. D. Skoromnik}
\ead{olegskor@gmail.com}

\author[label4]{V. G. Baryshevsky}

\author[label5]{A. P. Ulyanenkov}

\author[label1,label2,label3]{I. D. Feranchuk\corref{cor1}}
\ead{ilya.feranchuk@tdt.edu.vn}
\cortext[cor1]{Corresponding author}

\address[label6]{Max Planck Institute for Nuclear Physics, Saupfercheckweg 1, 69117 Heidelberg, Germany}

\address[label4]{Institute for Nuclear Problems, Belarusian State University, 4 Nezavisimosty Ave., 220030 Minsk, Belarus}

\address[label5]{Atomicus GmbH, Schoemperlen Str. 12a, 76185 Karlsruhe, Germany}

\address[label1]{Atomic Molecular and Optical Physics Research Group, Ton Duc Thang University, 19 Nguyen Huu Tho Str., Tan Phong Ward, District 7, Ho Chi Minh City, Vietnam}
\address[label2]{Faculty of Applied Sciences, Ton Duc Thang University, 19 Nguyen Huu Tho Str., Tan Phong Ward, District 7, Ho Chi Minh City, Vietnam}
\address[label3]{Belarusian State University, 4 Nezavisimosty Ave., 220030, Minsk, Belarus}

\begin{abstract}
  Parametric X-ray radiation (PXR) from relativistic electrons moving in a crystal along the crystal-vacuum interface is considered. In this geometry the emission of photons is happening in the regime of extremely asymmetric diffraction (EAD). In the EAD case the whole crystal length contributes to the formation of X-ray radiation opposed to Laue and Bragg geometries, where the emission intensity is defined by the X-ray absorption length. We demonstrate that this phenomenon should be described within the dynamical theory of diffraction and predict a radical increase of the PXR intensity. In particular, under realistic electron-beam parameters, an increase of two orders of magnitude in PXR-EAD intensity can be obtained in comparison with conventional experimental geometries of PXR. In addition we discuss in details the experimental feasibility of the detection of PXR-EAD.
\end{abstract}


\end{frontmatter}


\section{Introduction}
\label{sec:introduction}

Parametric X-ray radiation (PXR) occurs when a charg\-ed particle moves uniformly in a periodic medium \cite{PXR_Book_Feranchuk,J.Phys.France1985.46.1981} and possesses unique features such as high brightness, narrow spectral width and the possibility of tuning the X-ray frequency simply by rotating a crystal target. Moreover, PXR is emitted under a large angle with respect to the particle velocity and its brilliance is competitive with other X-ray sources, as already demonstrated experimentally \cite{HAYAKAWA200532}. Consequently, all these properties make it a suitable candidate for the development of novel-laboratory-compact X-ray sources with high brightness and tunable, quasi-monochromatic frequency.

There has been a lot of experimental research in this field \cite{PXR_Book_Feranchuk,PhysRevLett.79.2462,Brenzinger1997,PhysRevLett.79.4389,PhysRevLett.84.270,WAGNER20083893,Lauth2006,ALEXEYEV20162892,PUGACHOV200355,Aleinik2004,HAYAKAWA2006102,TAKABAYASHI201278,TAKABAYASHI201779,rullhusen1998novel} and at present an effort is made toward increasing the intensity of the PXR source. For example, the choice of the materials of the target was analyzed in Ref.~\cite{SONES200522}. In Ref.~\cite{AHMADI201378} it was demonstrated that under condition of anomalous absorption (the Borrmann effect) the PXR intensity is slightly increasing.

At the same time, in the majority of conventional experiments with PXR an electron beam is incident on a crystal under a large angle to its surface, i.e., transition geometry. In this situation according to kinematic model of diffraction \cite{J.Phys.France1985.46.1981} the PXR intensity is proportional to the smallest of either crystal $L$ or X-ray absorption $L_{\mathrm{abs}}$ lengths. In the X-ray frequency range $L_{\mathrm{abs}}\sim 10^{-2}$~cm and therefore in most cases $L_{\mathrm{abs}}\ll L$. For this reason, only a small part of the electron trajectory contributes to the formation of PXR.

As was mentioned above, PXR is emitted under a large angle with respect to the electron velocity, which makes it feasible to change the geometry of an experiment in a way such that the entire crystal length will contribute to the formation of PXR. Accordingly, this will lead to the increase of the total number of quanta in the PXR peak.
\begin{figure*}[t]
  \centering
  \includegraphics[width=0.42\textwidth]{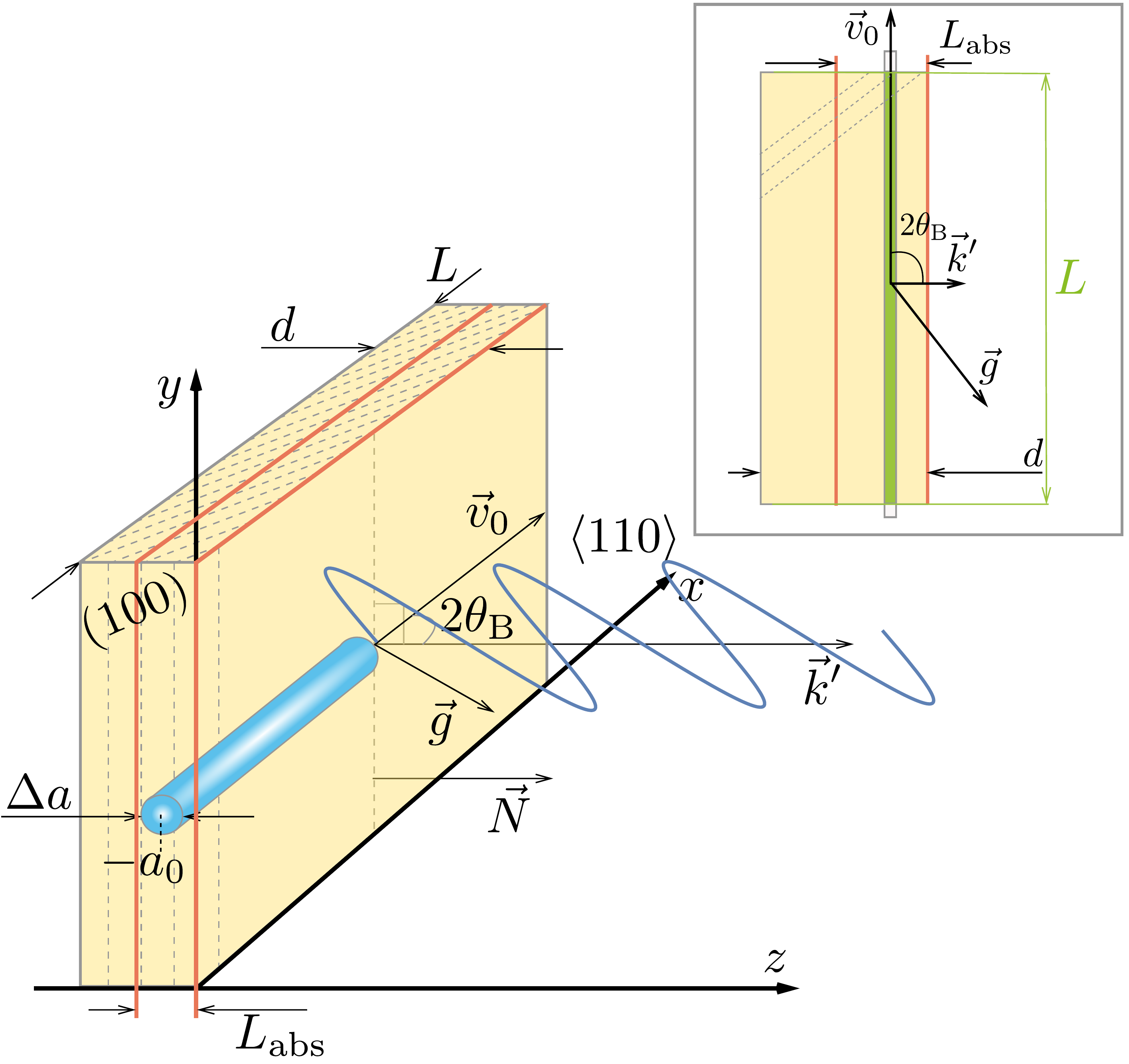}
  \includegraphics[width=0.52\textwidth]{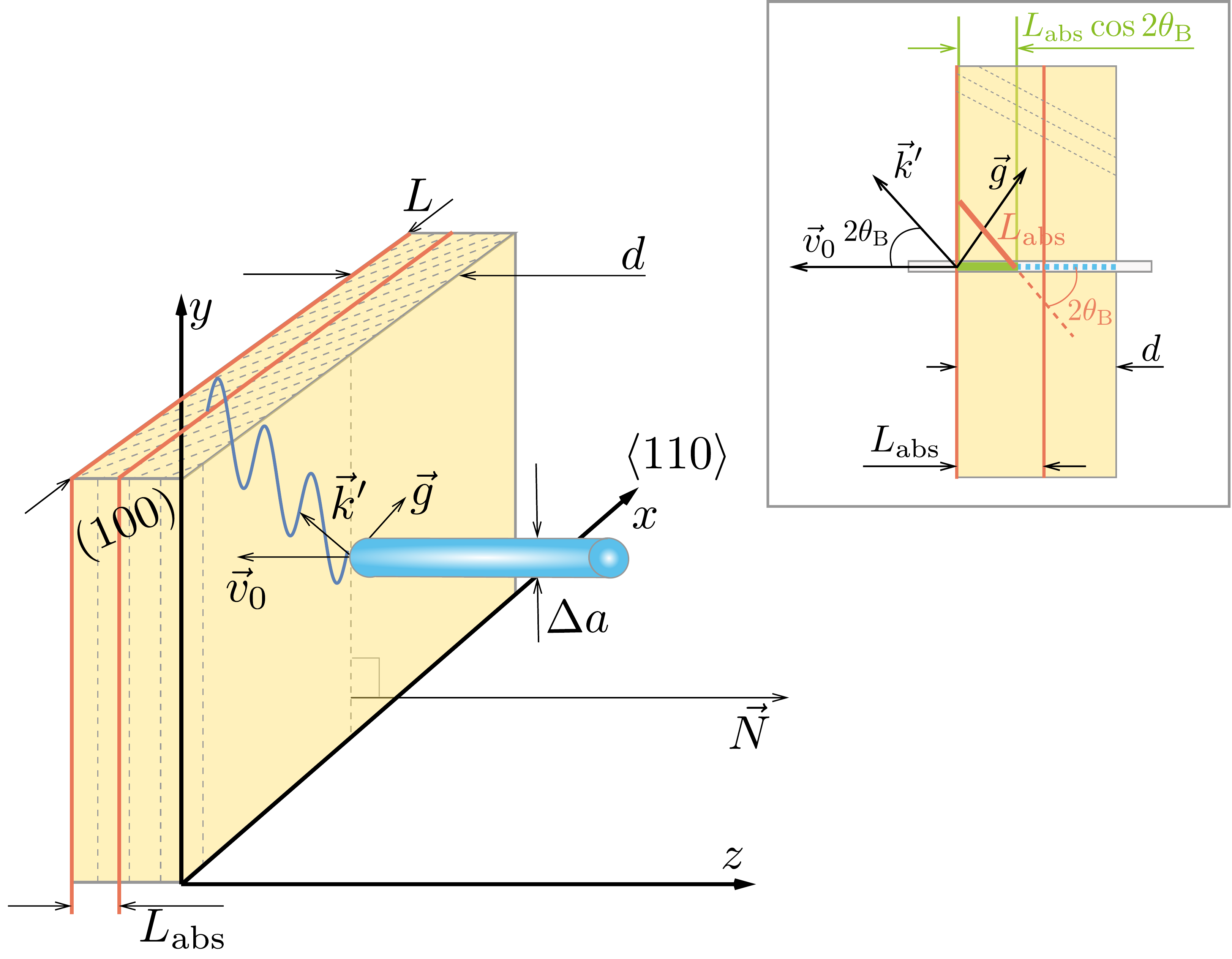}
  \caption{(Color online) Left pane: Grazing geometry of PXR-EAD. An electron beam propagates with velocity $\vec v_{0}$ along the $\langle110\rangle$ axis in a crystal parallel to the crystal-vacuum interface in a layer, whose thickness is smaller than $L_{\mathrm{abs}}$. The emitted radiation exits from a crystal in the direction $\vec k' = \omega \vec v_{0}/v_{0}^{2}+\vec g$ and is not absorbed. Right pane: Conventional transition geometry of PXR. An electron beam is incident on a crystal surface under a large angle. The propagation length of the emitted X-ray radiation is larger than the absorption length $L_{\mathrm{abs}}$. Consequently, only a small part of the electron trajectory $L_{\mathrm{abs}}\cos{2\theta_{\mathrm{B}}}$ contributes to the formation of the X-ray radiation, which leads to the decrease of the PXR intensity.}\label{fig:1}
\end{figure*}

In the first experiment of the detection of PXR \cite{pak1985experimental} the grazing geometry was used, when an electron beam was moving in a short crystal in a thin layer parallel to the crystal-vacuum interface and the emission occurred under a large angle with respect to the crystal surface. The first theoretical estimations were performed within the framework of the dynamical theory of diffraction under the condition of extremely asymmetric diffraction (EAD) \cite{BARYSHEVSKY1986306} and it was demonstrated that the whole crystal length may contribute to the formation of PXR, despite the condition $L_{\mathrm{abs}} < L$. Later, an analogous PXR geometry was discussed in Ref.~\cite{NASONOV200696,BLAZHEVICH20083777} and the increase of the PXR intensity was observed \cite{Eliseev2009}.

However, in \cite{BARYSHEVSKY1986306,NASONOV200696,Eliseev2009} the detailed analysis of the optimal conditions under which the PXR-intensity increase takes place has not been performed. In this work we fill this gap and provide a comprehensive theoretical analysis and show the experimental feasibility of the observation of PXR-EAD. Moreover, as will be shown below, we predict the PXR-EAD intensity being two orders of magnitude larger than the one observed by conventional transition geometries. Quantitative estimations will be provided for the parameters of the electron beam of Mainz Microtron MAMI, where one of the most detailed analysis of the PXR spectrum was performed \cite{Brenzinger1997}.

\section{Qualitative consideration}
\label{sec:qual-cons}

In order to discuss the qualitative characteristics of PXR-EAD we assume that a monocrystal plate of a thickness $d$ and a length $L$ is used as a target. In addition we consider that two realistic conditions $L\gg d$ and $L_{\mathrm{abs}} < d$ are also fulfilled. In Fig.~\ref{fig:1} the electron trajectories and tracks of emitted photons are plotted for two possible geometries of an experiment, namely the transition geometry (Laue case Fig.~\ref{fig:1} Right pane) and the grazing geometry of PXR-EAD (Fig.~\ref{fig:1} Left pane).

As was demonstrated in many works \cite{PXR_Book_Feranchuk,ter1972high} the formation of PXR is caused by the vanishing coherence length \cite{J.Phys.France1983.44.913}, as it takes place in the case of Cherenkov radiation. This means that all photons emitted along the electron trajectory have equal phase and are coherent. However, only photons, which are not absorbed in a crystal contribute to the detectable PXR peak. As follows from Fig.~\ref{fig:1} in the case of transition geometry the photons are emitted only on the part of an electron trajectory, which has the length $L_{\mathrm{abs}}\cos{2\theta_{\mathrm{B}}}$. Here $\theta_{\mathrm{B}}$ is the angle between the electron velocity $\vec v$ and the crystallographic planes, due to which the PXR peak is formed. This PXR peak is located under the angle $2\theta_{\mathrm{B}}$ with respect to $\vec v$ \cite{PXR_Book_Feranchuk}. As a result, the total number of quanta emitted in the case of transition geometry can be estimated as
 \begin{align}
   N_{\mathrm{PXR}} = Q_{\mathrm{PXR}}  L_{\mathrm{abs}}\cos{2 \theta_{\mathrm{B}}},\label{eq:1}
 \end{align}
where $Q_{\mathrm{PXR}}$ defines the number of photons emitted from the unit length of the electron trajectory. Its value can be estimated within the kinematic theory of diffraction \cite{J.Phys.France1985.46.1981}. For our qualitative analysis it is sufficient to know that $Q_{\mathrm{PXR}}$ is independent of the crystal length under the condition $L_{\mathrm{abs}} < d$.

Returning to the grazing geometry of PXR-EAD we observe that the absorption does not occur (Fig.~\ref{fig:1}). Let an electron beam with a transverse size $\Delta a$, an angular spread $\Delta \theta_{\mathrm{e}}$ and a natural emittance (not normalized) $\epsilon = \Delta a \Delta\theta_{\mathrm{e}}$ propagates in a crystal parallel to the crystal-vacuum interface. We denote as $\vec N$ the normal to the crystal surface. We also assume that the central part of the beam has a coordinate $z_{0} = -a_{0}$, $a_{0}<L_{\mathrm{abs}}$ and its velocity $\vec v_{0}$ is perpendicular to $\vec N$, viz. $\vec N \cdot \vec v_{0} = 0$. Finally, we consider that the PXR-EAD photons are emitted along $\vec N$. This geometry coincides with the experimental conditions of Ref.~\cite{pak1985experimental}. In this situation, all photons emitted from the whole electron trajectory $L$ are not absorbed, contribute to the formation of PXR-EAD and as will be shown below the Cherenkov condition is fulfilled. Consequently, we can write analogously to Eq.~(\ref{eq:1}) for the total number of emitted quanta of PXR-EAD
\begin{align}
  N_{\mathrm{PXR-EAD}} = Q_{\mathrm{PXR-EAD}}L.\label{eq:2}
\end{align}
\begin{figure}[t]
  \centering
  \includegraphics[width=\columnwidth]{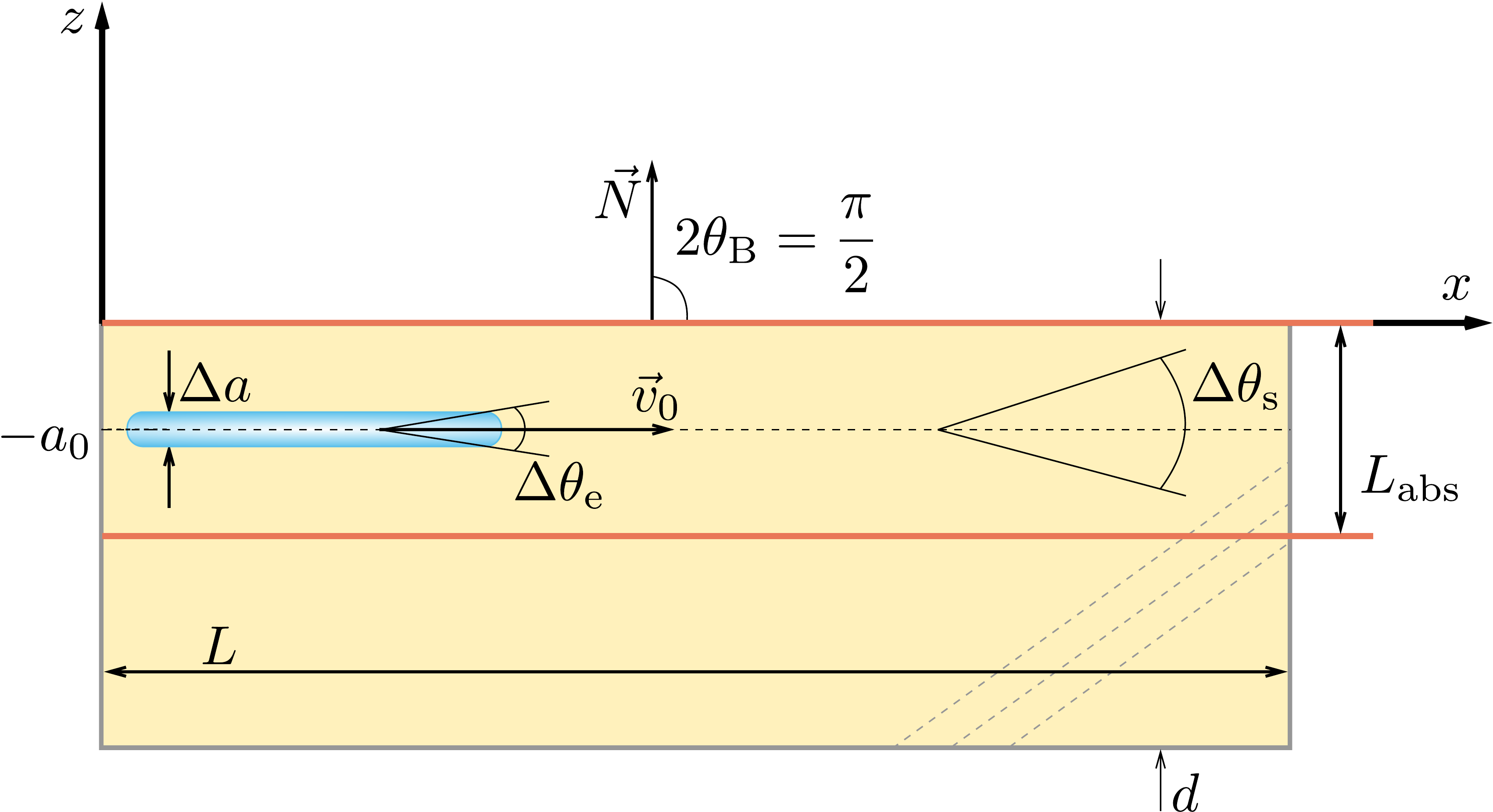}
  \caption{(Color online) The grazing geometry of PXR-EAD. An electron beam has an angular spread $\Delta\theta_{\mathrm{e}}$ and a transversal width $\Delta a$. Its central part has coordinate $z_{0} = -a_{0}$. $\Delta\theta_{\mathrm{s}}$ and $\vec N$ denote the electron scattering angle and the normal to the crystal surface respectively. $2\theta_{\mathrm{B}}$ is the angle under which radiation is emitted.}\label{fig:2}
\end{figure}

The exact value $Q_{\mathrm{PXR-EAD}}$ will be determined below. Here we only notice that its magnitude is comparable with $Q_{\mathrm{PXR}}$, i.e., $Q_{\mathrm{PXR}} \approx Q_{\mathrm{PXR-EAD}}$. Consequently, we can define a parameter $\xi$, which characterizes the increase of the intensity of PXR-EAD with respect to the intensity of PXR in the ideal case $\Delta a = \Delta \theta_{\mathrm{e}} = 0$
\begin{align}
  \xi = \frac{N_{\mathrm{PXR-EAD}}}{N_{\mathrm{PXR}}} \approx \frac{L}{L_{\mathrm{abs}}\cos{2\theta_{\mathrm{B}}}}.\label{eq:3}
\end{align}

However, the experimentally available electron beams impose constraints on the upper value of the parameter $\xi$. Indeed, in order the condition $a_{0}<L_{\mathrm{abs}}$ to be fulfilled the transverse width and the angular spread should satisfy inequalities (see Fig.~\ref{fig:2})
\begin{equation}
  \begin{aligned}
    &\Delta a < L_{\mathrm{abs}},
    \\
    &L \le \frac{\Delta a}{\Delta\theta_{\mathrm{e}}} \le \frac{L_{\mathrm{abs}}^{2}}{\epsilon},
  \end{aligned}\label{eq:4}
\end{equation}
which limit the actual value of the parameter $\xi$
\begin{align}
  \xi \le \frac{L_{\mathrm{abs}}}{\epsilon\cos{2\theta_{\mathrm{B}}}}.\label{eq:5}
\end{align}

Let us investigate the maximal value of the parameter $\xi$ from the inequality (\ref{eq:5}) within the experimental conditions of Ref.~\cite{Brenzinger1997}. For PXR from the crystalline planes (220) of a silicone crystal the following values of parameters were employed \cite{Brenzinger1997}
\begin{equation}
  \begin{aligned}
    \theta_{\mathrm{B}} &= 22.5^{\circ}, \quad \hbar\omega = 8.3\text{ keV},
    \\
    L_{\mathrm{abs}} &= 9.0\times 10^{-3}\text{ cm}, \quad \epsilon = 10^{-6}\ \mathrm{cm}\times\mathrm{rad}.
  \end{aligned}\label{eq:6}
\end{equation}

The estimation of the maximal value of the parameter $\xi$ Eq.~(\ref{eq:5}) with the help of Eq.~(\ref{eq:6}) yields
\begin{align}
  \xi_{\mathrm{max}} = 1.3\times 10^{4}.\label{eq:7}
\end{align}

However, in this case the crystal length $L = 81$ cm and, consequently, for the more realistic experimentally available crystals of the length $L\sim1$ cm, the actual value of the parameter $\xi$ via Eq.~(\ref{eq:3}) is given as $\xi\sim 10^{2}$.

Another restriction on the parameter $\xi$ follows from the multiple scattering of particles, which also withdraws electrons from the layer of the thickness $L_{\mathrm{abs}}$ (see Fig.~\ref{fig:2}). In this case we can estimate the mean square of the scattering angle according to Ref.~\cite{ter1972high}
\begin{align}
  \theta_{\mathrm{s}}^{2} = \left(\frac{E_{k}}{E}\right)^{2} \frac{L}{L_{\mathrm{R}}},\label{eq:8}
\end{align}
where $E$ is the energy of the electron measured in MeV, $E_{k}\approx 21$ MeV and $L_{\mathrm{R}}$ is the radiation length. (see also Ref.~\cite{TABRIZI20167})

Consequently, the following inequalities should be satisfied (see Fig.~\ref{fig:2})
\begin{align}
  &L\theta_{\mathrm{s}} < L_{\mathrm{abs}}\quad \Rightarrow\quad L < \left(\frac{E}{E_{k}}\right)^{2/3}(L_{\mathrm{abs}}^2 L_{\mathrm{R}})^{1/3} \quad\Rightarrow \nonumber
  \\
  &\xi < \left(\frac{E}{E_{k}}\right)^{2/3}\left(\frac{L_{\mathrm{R}}}{L_{\mathrm{abs}}}\right)^{1/3}. \label{eq:9}
\end{align}

For silicon the radiation length \cite{ter1972high} $L_{\mathrm{R}} \approx 9.6$ cm. The typical energy of the electron beam on MAMI facility \cite{Brenzinger1997,Lauth2006} $E = 900$ MeV. For this reason, for the realistic crystal length $L=1$ cm the multiple electron scattering does not prevent the parameter $\xi$ to reach $\xi\approx 10^{2}$.

Concluding, our qualitative considerations indicate that the optimization of the geometry of an experiment, namely the change from the transition to the grazing geometry, will provide the two orders of magnitude increase of the peak PXR intensity on the currently available experimental facilities.

 \section{Spectral--angular distribution and integral intensity of PXR-EAD}
\label{sec:spectr-angul-distr}

In the previous section we have introduced two quantities $Q_{\mathrm{PXR}}$ and $Q_{\mathrm{PXR-EAD}}$, which characterize the number of quanta emitted from the unit length of the particle trajectory. In this section we will employ the dynamical theory of diffraction and will determine the exact expression for $Q_{\mathrm{PXR-EAD}}$, which will validate the qualitative estimations given above.

Our derivation will be based on the approach developed in Refs.~\cite{PXR_Book_Feranchuk,J.Phys.France1983.44.913,baryshevsky2012high,doi:10.1143/JPSJ.69.3462}, when the solutions of the homogeneous Maxwell equation are used to calculate the number of emitted photons. It should be noted that the application of the solution of the homogeneous Maxwell equations instead of the solution of the inhomogeneous ones, as for example in \cite{NASONOV2005367}, significantly simplifies the analysis of the radiation problems and enables one to take into account multiple electron scattering \cite{PXR_Book_Feranchuk}. The number of photons with polarization $s =1,2$ that are emitted within a spectral interval $[\omega,\omega+d\omega]$ and a solid angle $d\Omega$ along a unit vector $\vec n$ by an electron of a charge $e_0$, which moves along a trajectory $\vec r (t)$ with a velocity $\vec v (t) = d \vec r (t)/dt$ is equal to
\begin{align}
\frac{\partial^2 N_{\vec n,\omega s}}{\partial \omega  \partial \Omega} = \frac{e_0^2  \omega }{4 \pi^2 \hbar c^3}  \left|\int \vec E_{\vec k' s}^{(-)\ast} (\vec r (t),\omega) \vec v (t) e^{i\omega t} dt\right|^2.\label{eq:10}
\end{align}
Here $\vec E_{\vec k' s}^{(-) } (\vec r,\omega)$ is an electric field strength of an electromagnetic wave with a polarization $\vec e_s$ corresponding to the solution of the homogeneous Maxwell equations. The integration in Eq.~(\ref{eq:10}) is carried out over the whole particle trajectory.

The electric field in Eq.~(\ref{eq:10}) satisfies an asymptotic boundary condition when $r\rightarrow\infty$ corresponding to the superposition of a plane wave and an ingoing spherical wave
\begin{align}
\vec E_{\vec k' s}^{(-)} (\vec r,\omega) \approx \vec e_s e^{i\vec k'\cdot\vec r} + \mathrm{const}\cdot \frac{ e^{-i k' r} }{r}, \label{eq:11}
\end{align}
which is different from the conventional relation for the solution $\vec E_{\vec k' s}^{(+)} $ of the Maxwell equations, which contains as $r\rightarrow\infty$ an outgoing spherical wave. The field $\vec E_{\vec k' s}^{(-)}$ is related to the field $\vec E_{\vec k' s}^{(+)}$ with the following relation
\begin{align}
\vec E_{\vec k' s}^{(-)\ast} = \vec E_{\vec k s}^{(+)}, \quad  \vec k = - \vec k'\label{eq:12}
\end{align}
that is an analog of the well known `reciprocity theorem' in classical optics \cite{born2013principles}. For this reason, instead of Eq.~(\ref{eq:10}) we will employ the modified expression
\begin{align}
  \frac{\partial^2 N_{\vec n,\omega s}}{\partial \omega  \partial \Omega} = \frac{e_0^2  \omega }{4 \pi^2 \hbar c^3}  \left|\int \vec E_{\vec k s}^{(+)} (\vec r (t),\omega) \vec v (t) e^{i\omega t} dt\right|^2,\label{eq:13}
\end{align}
which contains the field amplitude $\vec E^{(+)}_{\vec ks}$.

Consequently, in order to calculate the number of emitted quanta with Eq.~(\ref{eq:13}) we firstly need to determine the solution of the homogeneous Maxwell equations $\vec E_{\vec k s}^{(+)}$. For this we will employ the dynamical theory of diffraction. As was mentioned above, for the observation of PXR-EAD the crystal thickness should be larger that the absorption length. Under this realistic assumption we can employ the two wave approximation of the dynamical diffraction theory \cite{authier2001dynamical,benediktovich2013theoretical}. Within this framework two strong diffraction waves are excited. Let us separate out the scalar field amplitudes for incident $\vec E_{\vec k s}^{(+)} = \vec e_{s} E_{\vec ks}$ and diffracted $\vec E_{\vec k_{g} s}^{(+)} = \vec e_{1s} E_{\vec k_{g}s}$ waves. These amplitudes satisfy a set of algebraic equations \cite{PXR_Book_Feranchuk}:
\begin{equation}
  \begin{aligned}
    &\left(\frac{k^2}{k_0 ^2} -1 - \chi_0 \right) E_{\vec k s}  - c_s \chi_{-\vec g} E_{\vec k_g s} = 0,
    \\
    &\left(\frac{k_g^2}{k_0^2} -1 - \chi_0\right)   E_{\vec k_g s}  - c_s \chi_{\vec g}    E_{\vec k s} =0,
  \end{aligned}\label{eq:14}
\end{equation}
where $k_0 = \omega/c$, $\vec k_g = \vec k+ \vec g$, $\vec g$ is the reciprocal lattice vector, $\chi_0$ and $\chi_{\vec g}$ are the Fourier components of the crystal susceptibility $\chi(\vec r)$:
\begin{align}
  \chi(\vec r)= \sum_{\vec g} \chi_{\vec g} e^{i\vec g\cdot\vec r},\label{eq:15}
\end{align}
$c_s = 1 $  for $\sigma$ ($s=1$) and $c_s = \cos 2\theta_{\mathrm{B}}$ for $\pi$  ($s=2$) polarizations of the incident and diffracted waves respectively. In the following we denote the real and imaginary parts of the dielectric susceptibilities with a single and double prime respectively $\chi_{\vec g} = \chi'_{\vec g} + i \chi_{\vec g}''$. Moreover, it is well known \cite{authier2001dynamical,benediktovich2013theoretical} that waves of different polarizations propagate independently with an accuracy up to $|\chi_{0}|^{2}$.
\begin{figure*}[t]
  \centering
  \includegraphics[width=\columnwidth]{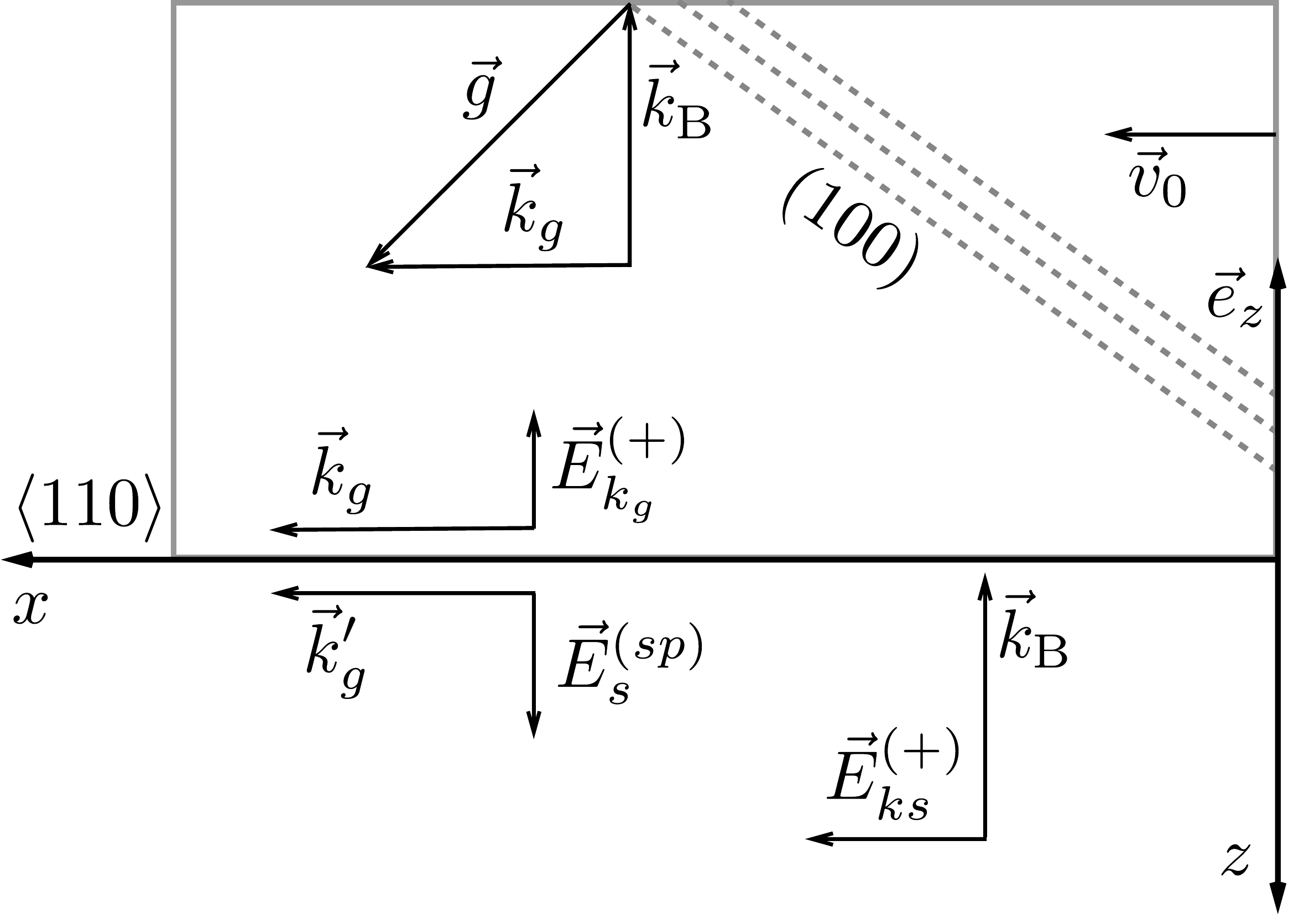}
  \includegraphics[width=\columnwidth]{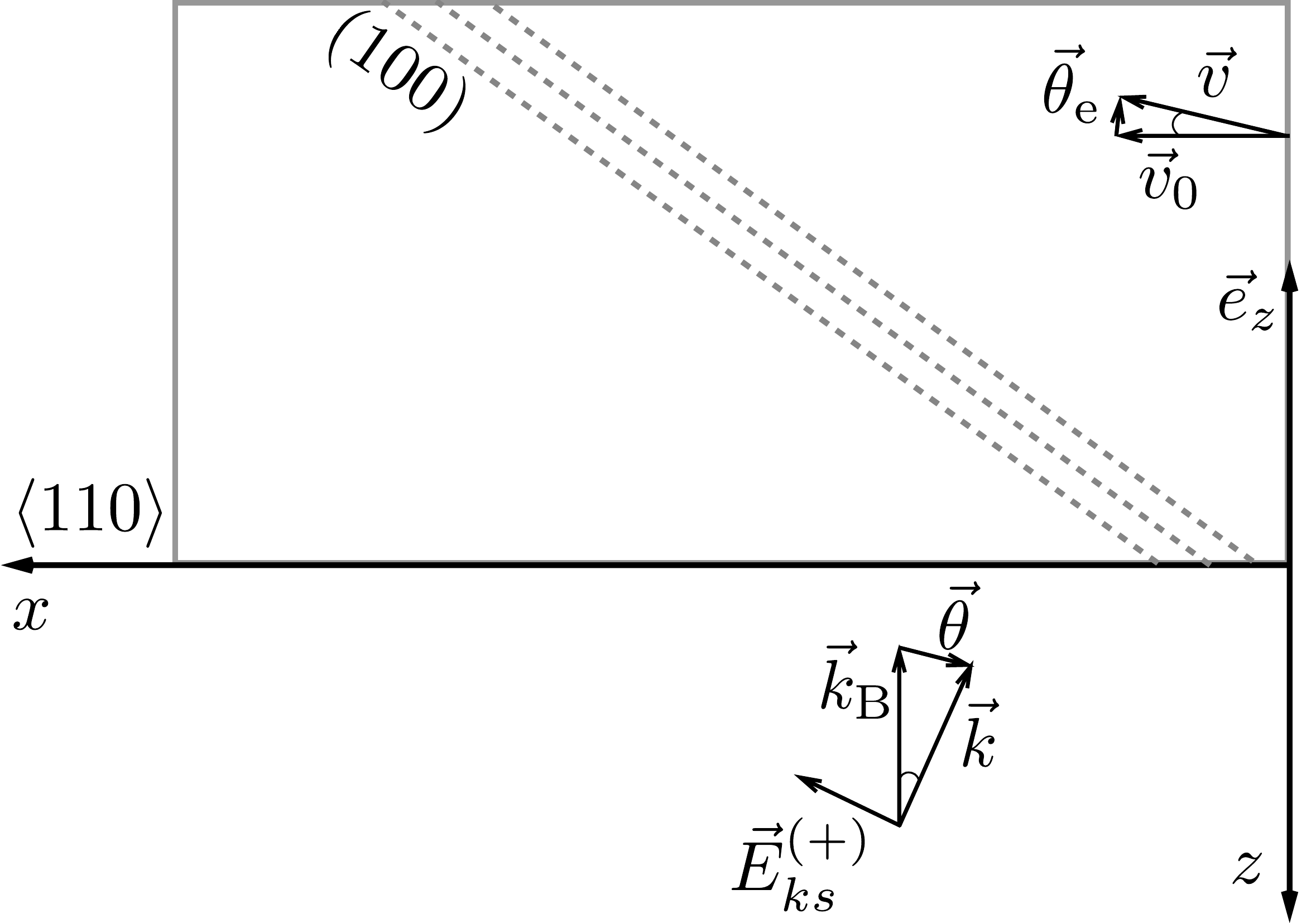}
  \caption{Left pane: The grazing geometry of PXR-EAD in the ideal case, when the Wulff–Bragg's condition for the emitted photons is fulfilled and propagation directions of incident, diffracted and diffracted-reflected waves. The angle $\theta_{0}$ is the angle between $\vec v_{0}$ and the $x$-axis and in the ideal case $\theta_{0} = 0$. Right pane: Non ideal case. The deviation from the Wulff–Bragg's condition and the variation of the velocity of the center of the beam are described by the vectors $\vec \theta$ and $\vec\theta_{\mathrm{e}}$ respectively. In this pane the diffracted and diffracted-reflected waves are not shown. In both panes the incident wave $\vec E_{\vec k s}^{(+)}$ describes the emitted PXR field in accordance with reciprocity theorem Eq.~(\ref{eq:12}).}\label{fig:3}
\end{figure*}

The system of Eqs.~(\ref{eq:14}) is a system of linear uniform equations. Consequently, in order this system to be solvable its determinant should vanish. This provides us with a dispersion equation and allows us to determine the possible wave vectors in a crystal. In addition, we can find the relation between scalar amplitudes of incident $E_{\vec k s}$ and diffracted $E_{\vec k_g s}$ waves respectively
\begin{align}
  &\left(\frac{k^2}{k_0 ^2} -1 - \chi_0\right)\left(\frac{k_g^2}{k_0^2} -1 - \chi_0\right) - c_s^2\chi_{\vec g}\chi_{-\vec g} = 0, \label{eq:16}
  \\
  &E_{\vec k_g s} = V_{\vec k s} E_{\vec k s}, \quad V_{\vec k s} = \frac{(\frac{k^2}{k_0^2} -1 - \chi_0 )  }{c_s \chi_{-\vec g} }. \label{eq:17}
\end{align}

In Fig.~\ref{fig:3} we show the propagation directions of electromagnetic fields in vacuum and in crystal for our grazing geometry. This type of geometry corresponds to the EAD case (grazing exit) \cite{authier2001dynamical,benediktovich2013theoretical}. However, in vacuum one should take into account not only an incident wave $\vec E_{\vec k s}^{(0)}$, but also a diffracted wave $\vec E_{\vec k_g s}^{(sp)}$, which is specular reflected 
\begin{align}
  \vec E_{\vec k s}^{(0)} &= \vec e_s e^{i \vec k\cdot \vec r}, \nonumber
  \\
  \vec E_{\vec k_g s}^{(sp)} &= \vec e_{1s} E_{\vec k_g s}^{(sp)} e^{i (\vec k_{\|} + \vec g_{\|})\cdot\vec r} e^{i  k'_{gz}  z}, \label{eq:18}
  \\
  k'_{gz} &=  \sqrt{k_0^2 - (\vec k_{\|} + \vec g_{\|})^2 }. \nonumber
\end{align}

For a given polarization of an electromagnetic field in a crystal there exist four solutions of the dispersion Eq.~(\ref{eq:16}). However, two out of four these solutions are unphysical, as the corresponding values of $k$ lead to the exponentially growing solutions for electromagnetic waves inside a crystal. Consequently, we need to take into account only two electromagnetic waves inside the crystal with a positive imaginary part of $k_{z}$, which defines the $z$ component of the wave vector $\vec k$ in a medium \cite{authier2001dynamical,benediktovich2013theoretical}. In addition, due to the boundary conditions on the crystal-vacuum interface the in-plane components $\vec k_{\|}$ in vacuum and in the crystal are equal. For this reason the correction to the wave vector is defined through the change of the projection of the wave vector on the normal $\vec N$ to the surface \cite{PXR_Book_Feranchuk}
\begin{align}
  \vec k_{s \mu} = \vec k  - k_0 \epsilon_{s \mu}\vec N, \quad \mu = 1,2. \label{eq:19}
\end{align}
Here $\vec k = k_0 \vec n$ and $\vec n$ is a unit vector in the direction of an incident wave in vacuum. In the considered geometry the value $\nu_{0} = (\vec k\cdot\vec N)/k_0 \approx - 1$. At the same time, the wave vector of the diffracted wave $\vec k_g = \vec k + \vec g$ is directed parallel to the crystal surface and, consequently, $|\nu_{g}| = |(\vec k_g \cdot \vec N)/k_0| \ll 1$. 

As a result, $\epsilon_{s \mu}$ in Eq.~(\ref{eq:19}) are defined as two solutions of the following cubic equation
\begin{align}
  &- 2 \nu_{0} \epsilon^3 + (4\nu_{0}\nu_{g} - \chi_0) \epsilon^2+ 2 \nu_{0} (\chi_0     - \alpha_{\mathrm{B}})\epsilon + \chi_0^2 \nonumber
  \\
  &\mspace{45mu}- \chi_0 \alpha_{\mathrm{B}} - c_s^2\chi_{\vec g}\chi_{-\vec g} = 0, \label{eq:20}
\end{align}
where we disregarded the part from the specular wave, since its amplitude is small in the considered grazing geometry \cite{authier2001dynamical,benediktovich2013theoretical}. The quantity $\alpha_{\mathrm{B}}$ in Eq.~(\ref{eq:20}) defines the deviation from the Wulff–Bragg's condition
\begin{align*}
  \alpha_{\mathrm{B}} = \frac{k^2 - k_g^2}{k_0^2} = - \frac{2 \vec k \cdot\vec g + g^2}{k_0^2}.
\end{align*}

We note here that the vector $\vec k_{\mathrm{B}}$ in Fig.~\ref{fig:3} corresponds to the condition $\alpha_{\mathrm{B}} = 0$. In addition, we highlight here that in accordance with the above mentioned reciprocity theorem Eq.~(\ref{eq:12}) the incident wave with the wave vector $\vec k$ describes the PXR wave, which is emitted in the observation direction $\vec k' = -\vec k$.

In the general case the solutions of the cubic Eq.~(\ref{eq:20}) are given by cumbersome analytical expressions. However, simple analytical approximate solutions can be found if one considers the following fact \cite{PSSA:PSSA2210710211}. It is well known that the angular spread in the PXR peak is defined via the parameter $\gamma^{-1} = mc^2/E \sim \sqrt{|\chi_0|}$. Consequently, the conditions $|\nu_{0}|\approx 1$ and $|\nu_{g}| \approx \sqrt{|\chi_0|} \gg |\chi_0|$ are satisfied. Within this approximation the desired roots are
\begin{align}
  \epsilon_{1s} &= - \frac{\chi_0}{2 \nu_{0}} + \frac{c_s^2\chi_{\vec g}\chi_{-\vec g}}{2\nu_{0} (\alpha_{\mathrm{B}} + \chi_0) },\label{eq:21}
  \\
  \epsilon_{2s} &= \nu_{g}  + \sqrt{\nu_{g}^2 + \alpha_{\mathrm{B}} + \chi_0},\label{eq:22} 
  \\
  |\epsilon_{2s}| &\sim \sqrt{|\chi_0|} \gg  |\epsilon_{1s}| \sim  |\chi_0|, \nonumber
  \\
  \epsilon_{2s}'' &\sim \frac{\chi_0''}{\sqrt{|\chi_0|}} \gg   \epsilon_{1s}'' \sim   \chi_0''. \label{eq:23}
\end{align}

Hence, the desired solutions of the Maxwell equations in the interval $0<x<L$ in crystal ($z<0$) and in vacuum ($z>0$) are represented as the following superposition
\begin{align}
  \vec E_{\vec k s}^{(+)} &= \vec e_s e^{i \vec k \cdot\vec r} + \vec e_{1s} E_{  s}^{(sp)} e^{i (\vec k_{\|} + \vec g_{\|})  \cdot\vec r} e^{i  k'_{gz}  z}, \quad z < 0, \label{eq:24}
  \\
  \vec E_{\vec k s}^{(+)} &= e^{i \vec k\cdot \vec r}\sum_{\mu = 1,2}e^{-ik_0 z \epsilon_{\mu s}} ( \vec e_s E_{\mu s} + \vec e_{1s} E_{ g \mu s}  e^{i  \vec g\cdot \vec r}),\quad z > 0. \label{eq:25}
\end{align}

In order to determine the amplitudes of these waves one shall impose the continuity of the field on the crystal surface, which in the EAD case yields the system of equations \cite{authier2001dynamical,benediktovich2013theoretical}
\begin{equation}
  \left\{
    \begin{aligned}
      &E_{1 s} + E_{2 s} = 1,
      \\
      &E_{s}^{(sp)} = E_{g 1 s} + E_{g 2 s},
      \\
      &\nu_{g}' E_{  s}^{(sp)} = (\nu_{g} - \epsilon_{1 s}) E_{g 1 s} + (\nu_{g} - \epsilon_{2 s}) E_{g 2 s},
    \end{aligned}
\right.\label{eq:26}
\end{equation}
where $\nu_{g}' = (\vec k'_{g} \cdot\vec N) / k_{0}$.

Finally, the solutions of Eqs.~(\ref{eq:26}) are easily obtained
\begin{equation}
  \left\{
    \begin{aligned}
      &E_{1 s}  = \frac{(2 \epsilon_{2 s} + \nu_{0}) (\nu_{g}' - \nu_{g} +\epsilon_{2 s}) }{(\epsilon_{2 s} - \epsilon_{1 s})[2\nu_{0} (\nu_{g}' - \nu_{g}  + \epsilon_{2 s} + \epsilon_{1 s}) + \chi_0 ]},
      \\
      &E_{2 s} = 1 - E_{1 s},
      \\
      &E_{ g \mu s} = - \frac{2\nu_{0} \epsilon_{\mu s} + \chi_0 }{c_s \chi_{-g}} E_{ \mu s}.
    \end{aligned}
\right.\label{eq:27}
\end{equation}

Proceeding with the calculation of the number of emitted quanta Eq.~(\ref{eq:13}) we notice that according to its definition the PXR radiation corresponds to the uniform motion of a particle, i.e.,
\begin{align}
  \vec r(t) = \vec r_0 + \vec v t, \quad \vec r_0 = (0,0,z_0).\label{eq:28}
\end{align}
Here $\vec r_{0}$ is an electron coordinate in the crystal $(x,z)$ plane. Moreover, without the loss of generality we can consider $r_{0y} = 0$.

When an electron velocity $\vec v$ is directed perpendicular to the normal of the surface $\vec N$ ($\vec v\bot\vec N$) the radiation is formed by the field of the diffracted wave, which propagates along the electron velocity $\vec v$ ($\vec v \|(\vec k + \vec g)$). For this wave the Cherenkov condition $\vec v\cdot\vec k_{g} = \omega/c$ can be fulfilled \cite{J.Phys.France1983.44.913}. In the case when $z_{0}>0$, i.e. the particle travels above the crystal surface, the radiation is caused by the field $E_{s}^{(sp)}$. Accordingly, this contribution is exponentially suppressed $\sim \exp(-\gamma^{-1}k_{0}|z_{0}|)$, $\gamma = E/(mc^{2})$, as in the case of Cherenkov radiation \cite{ginzburg2013theoretical}. Consequently, in the X-ray frequency range the stringent requirements need to be imposed on the transverse size of the electron beam.

In the case when $z_{0}<0$ the emitted radiation is formed by the diffracted waves $E_{g1,2s}\sim \exp{(- \epsilon_{1,2s}'' k_0 |z_0|)}$ in a crystal. However, the estimation (\ref{eq:23}) for the dielectric susceptibility $\epsilon_{2s}''$ leads to $E_{g2s}\sim \exp{(-   k_0 |z_0|\chi_0''/\sqrt{|\chi_0|})}$, which is also exponentially suppressed for realistic transversal beam sizes.
\begin{figure*}[t]
  \centering
  \includegraphics[width=\columnwidth]{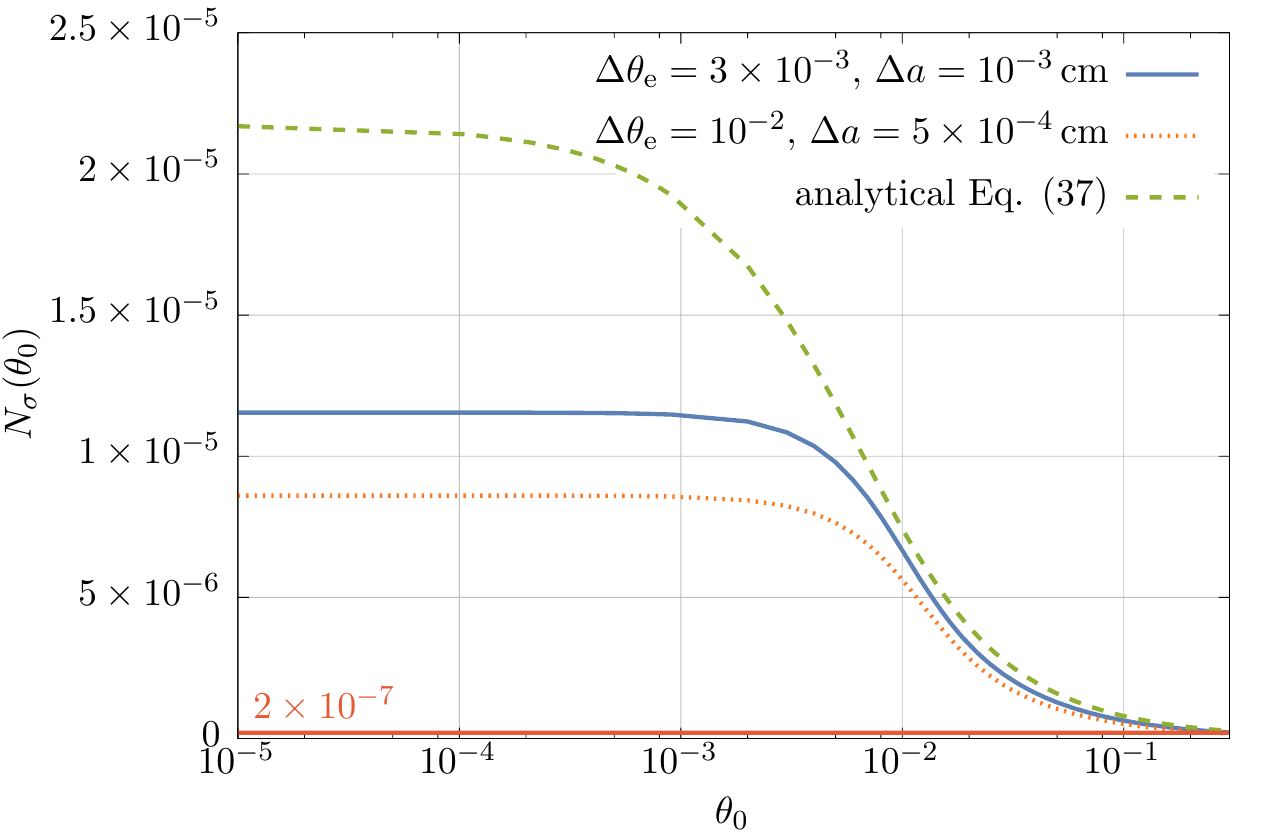}
  \includegraphics[width=\columnwidth]{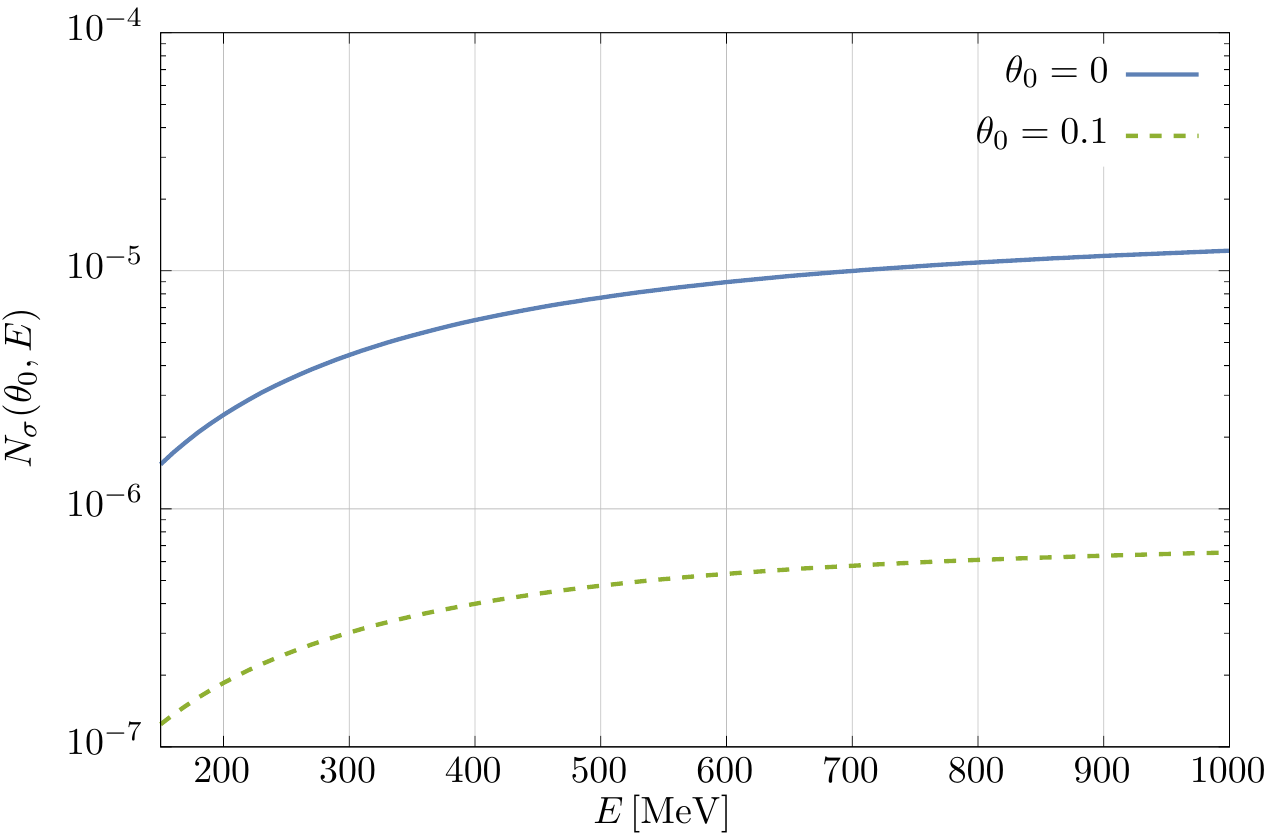}
  \caption{(Color online) Left pane: The dependence of the total number of emitted photons on the angle $\theta_{0}$, which defines the orientation of a crystal and an electron beam (See Fig.~\ref{fig:3}). The blue solid and orange dotted lines describe the weighted average with the Gaussian probability distribution. The green dashed line is the approximate analytical expression (\ref{eq:38}). The red solid line corresponds to the number of emitted quanta $2\times 10^{-7}$, when the electrons are entering the crystal at the angle $\theta_{0} \approx 0.31$, i.e., for the transition geometry. When the geometry is changed from the transition to the grazing one ($\theta_{0}\to0$) the number of photons exponentially increases. Right pane: The dependence of the number of emitted photons on the electron energy $E\,[\mathrm{MeV}]$. The blue solid line is the grazing geometry $\theta_{0} = 0$ and the green dashed line is the transition geometry $\theta_{0} = 0.1$. For both figures the parameters via Eq.~(\ref{eq:41}) were employed.}\label{fig:4}
\end{figure*}

As a result, we can perform integration over an electron trajectory in Eq.~(\ref{eq:13}) with $E_{g1s}$ only and find the following expression for the spectral-angular distribution of the PXR-EAD intensity
\begin{align}
  \frac{\partial^2 N_{\vec n,\omega s}}{\partial \omega  \partial \Omega} &= \frac{e_0^2  \omega }{4 \pi^2 \hbar c^5} (\vec e_{1s}\cdot\vec v)^2  | E_{ g 1 s} L_g (1 - e^{- i L/L_g})|^2 \nonumber
  \\
  &\mspace{45mu}\times e^{ -2 \epsilon_{1s}'' k_0 |z_0| },\label{eq:29}
\end{align}
where
\begin{align*}
  L_g &= \frac{c}{\omega - \vec v\cdot (\vec k + \vec g) + \vec v\cdot \vec N k_0\epsilon_{1s} } \equiv \frac{1}{k_0 q},
  \\
  q &= 1 - \frac{\vec v\cdot (\vec k + \vec g)}{\omega} + \frac{\vec v\cdot \vec N  \epsilon_{1s}}{c}.
\end{align*}
Here the quantity $L_{g}$ is the so called coherent length \cite{Galitsky1964} of the considered radiation, $k_{0}q$ is the longitudinal component of the wave vector, which defines the recoil exhibited by the electron in the emission process \cite{J.Phys.France1983.44.913}.

The approximation (\ref{eq:23}) also significantly simplifies the expression for the field amplitude $E_{g1s}$ in the expression (\ref{eq:29})
\begin{align}
  E_{g1s} = \frac{c_s \chi_g }{\alpha_{\mathrm{B}} + \chi_0}.\label{eq:30}
\end{align}

For our grazing geometry $\theta_B = \pi/4$, $c_{\sigma} = 1$, $c_{\pi} = \cos 2\theta_B \approx 0$. Consequently, the emitted radiation will be polarized in the plane perpendicular to the one defined by the vectors $\vec N$ and $\vec g$.

In the X-ray frequency range $k_0 L \gg 1$ and the analytical analysis of the number of emitted quanta Eq.~(\ref{eq:29}) can be performed with the help of the following asymptotic relation \cite{J.Phys.France1985.46.1981}
\begin{align}
  |L_g (1 - e^{- i L/L_g})|^2 \approx \frac{\pi}{k_0^2} \delta [\re q]  \frac{1 - e^{- 2 k_0 L |\im q|}}{|\im q|},\label{eq:31}
\end{align}
which is valid when  $|\im q| \ll |\re q|$. Here $\delta(x)$ is a delta function.

As follows from Fig.~\ref{fig:1} the spread of an electron velocity in the $(z,y)$ plane is small and the wave vector $\vec k = k_{0} \vec n$ is directed along $\vec e_{z} = -\vec N$ with a small angular spread in the plane $(x,y)$ (see Fig.~\ref{fig:3}). Then with the accuracy up to $O(|\vec\theta|^{3} \simeq \gamma^{-3})$ one can write
\begin{align}
  \vec n = \vec e_z \cos|\vec \theta| + \vec \theta, \quad  \vec v = v_0 (\vec e_x \cos|\vec \theta_{\mathrm{e}}| + \vec \theta_{\mathrm{e}}),\label{eq:32}
\end{align}
where $\vec \theta = (\theta_x, \theta_y)$ and $\vec \theta_{\mathrm{e}} = (\theta_{\mathrm{e}z}, \theta_{\mathrm{e}y})$. With the help of these notations we find for the parameter $\re q$
\begin{align}
  \re q \approx  1 -  \frac{\vec v\cdot \vec g }{\omega} - \frac{v_0}{c}(\theta_x + \theta_{\mathrm{e}z} + \vec \theta \cdot\vec \theta_{\mathrm{e}}) -  \theta_{\mathrm{e}z} \epsilon_{1s}'. \label{eq:33}
\end{align}

The Cherenkov condition $\re q = 0$ defines the frequency $\omega_{\mathrm{B}}$ of the emitted radiation in the PXR-EAD peak. With the accuracy $\gamma^{-2}$, $\omega_{\mathrm{B}}$ is equal to
\begin{align}
  \omega_{\mathrm{B}} = \frac{\vec v\cdot\vec g}{1 - (\theta_x + \theta_{\mathrm{e}z} + \vec \theta\cdot \vec \theta_{\mathrm{e}})} \approx \frac{g v_0}{\sqrt{2}} (1 + \theta_x ).\label{eq:34}
\end{align}

We pay attention here that with the considered accuracy the deviation of the electron velocity from the $x$-axis does not change the frequency of the emitted radiation. At the same time the deviation from the Wulff-Brag's condition defines the magnitude of the field $E_{g1s}$. With the above accuracy one finds for $\alpha_{\mathrm{B}}$ and $\im q$
\begin{align}
  \alpha_{\mathrm{B}} &= -2\left[ 1 - \frac{v_0}{c} \cos |\vec\theta| \cos |\vec\theta_{\mathrm{e}}| - \theta_y \theta_{\mathrm{e}y} - \theta_x \theta_{\mathrm{e}z}\right] \nonumber
  \\
  &\approx -[ \gamma^{-2} + (\theta_y - \theta_{\mathrm{e}y})^2 + (\theta_x -\theta_{\mathrm{e}z})^2 ],\label{eq:35}
  \\
  \im q &\approx  \frac{1}{2}\theta_{\mathrm{e}z}   \chi_0''\left[ 1 - \frac{|\chi_g|^2}{(\alpha_{\mathrm{B}} + \chi_0')^2}\right] \approx \frac{1}{2}\theta_{\mathrm{e}z}\chi_0''.\label{eq:36}
\end{align}
In Eq.~(\ref{eq:36}) we disregarded the second term in square brackets as the numerator in that fraction is approximately one order of magnitude smaller than the denominator \cite{StepanovXrayWebServer}.

Finally by plugging Eqs.~(\ref{eq:36}), (\ref{eq:35}) and (\ref{eq:33}) into Eq.~(\ref{eq:29}) and integrating out the frequency one finds a relatively simple angular distribution
\begin{align}
  \frac{\partial^2 N_{ \sigma}}{\partial \theta_x  \partial \theta_y} &= \frac{e_0^2    }{4 \pi \hbar c } \frac{(\theta_y - \theta_{\mathrm{e}y})^2 |\chi_g|^2 }{[\gamma^{-2} + (\theta_y - \theta_{\mathrm{e}y})^2 + (\theta_x -\theta_{\mathrm{e}z})^2 - \chi_0']^2}\times \nonumber
  \\
  &\times  \frac{(1 - e^{-   L k_0\chi_0''|\theta_{\mathrm{e}z}|} )}{\chi_0''|\theta_{\mathrm{e}z}|} e^{ -\chi_0'' k_0 |z_0| }\label{eq:37}
\end{align}
and all characteristics of the medium are evaluated for $\omega = \omega_{\mathrm{B}}$.

In order to determine the total number of emitted X-ray photons we will perform an angular integration in Eq.~(\ref{eq:37}) weighted with the Gaussian probability distributions of $\theta_{\mathrm{e}y,z}$ and $z_{0}$ due to the beam spread and multiple electron scattering. In addition, we perform a simple approximate analytical estimation.

For the numerical evaluation we consider that the angle $\theta_{\mathrm{e}z}$ is counted from the angle $\theta_{0}$ and introduce polar coordinates $\theta_{x} - \theta_{0} = \rho\cos\varphi$, $\theta_{y} = \rho\sin\varphi$. The integration is then performed in the range $\rho = [0, \theta_{\mathrm{D}}]$ and $\varphi = [0,2\pi]$, with $\theta_{\mathrm{D}}$ being the aperture of a detector. The result of this integration is then convoluted with the normalized Gaussian distribution $\exp[-(\theta_{\mathrm{e}z}^{2} + \theta_{\mathrm{e}y}^{2}) / (\theta_{\mathrm{s}}^{2} + \Delta\theta_{\mathrm{e}}^{2})] \times \exp[(|z_{0}| - a_{0})^{2} / \Delta a^{2}]$. We also note that the integration with respect to $|z_{0}|$ is performed in the range $[0,\infty)$.
\begin{figure}[t]
  \centering
  \includegraphics[width=\columnwidth]{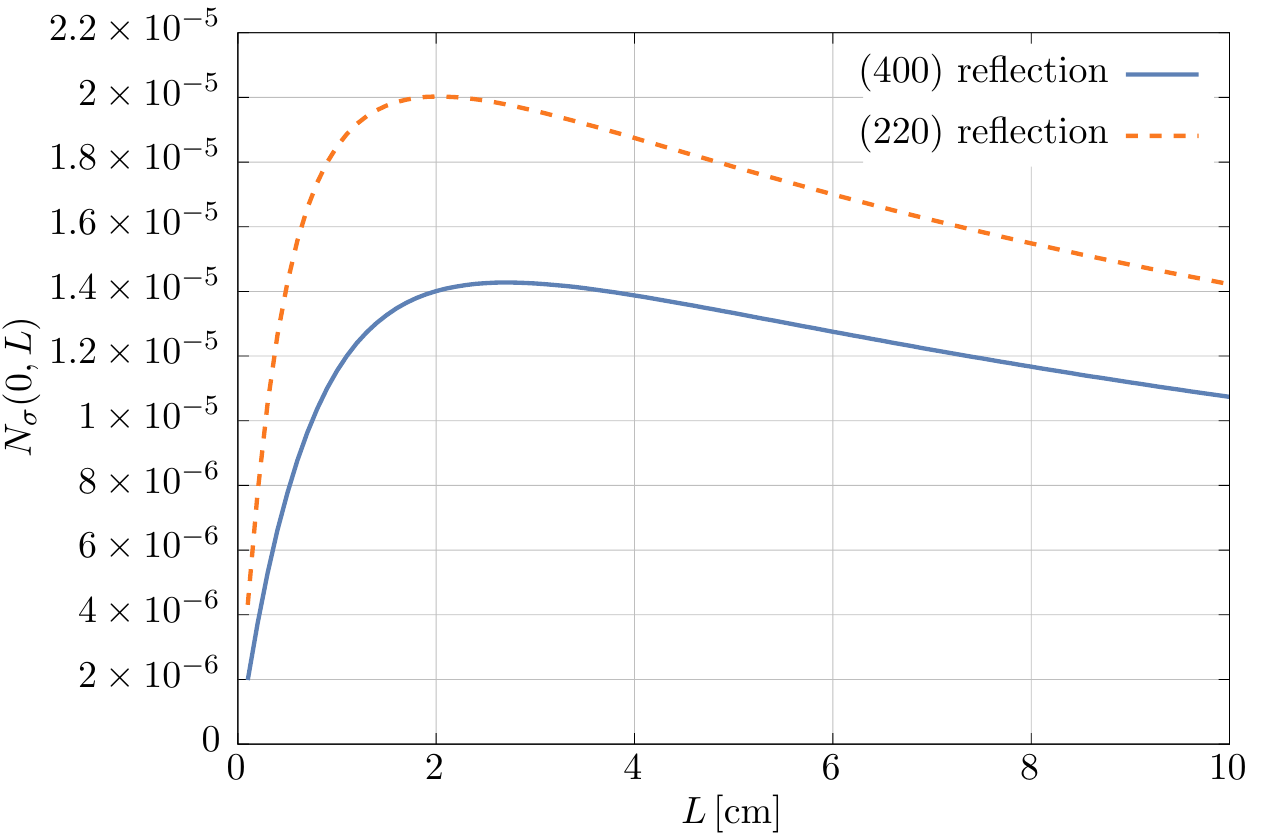}
  \caption{(Color online) The dependence of the number of emitted quanta on the crystal length $L\,[\mathrm{cm}]$, for the parameters $\theta_{0} = 0$ and $E = 900$ MeV. The blue solid line describes $(400)$ reflection. The dashed orange line is for $(220)$ reflection.}
  \label{fig:5}
\end{figure}

For the analytical approximation we assume that the electron beam satisfies the conditions (\ref{eq:4}) and (\ref{eq:9}), which were fulfilled during the experiment of Ref.~\cite{Brenzinger1997}. Moreover, we consider that the beam entrance angle is located near $\theta_{0}$ and $\theta_{\mathrm{e}z}\sim \theta_{0}$. During an experiment the angle $\theta_{0}$ can vary within the angle of total external reflection of the X-ray radiation with frequency $\omega_{\mathrm{B}}$, $\theta_0 \le \sqrt{|\chi_0(\omega_{\mathrm{B}})|}$ \cite{baryshevsky2012high}. The multiple electron scattering will be taken into account in analogy with the kinematic model of diffraction by substituting $\langle\theta_{\mathrm{e}}^{2}\rangle \to \theta_{\mathrm{s}}^{2}$, Eq.~(\ref{eq:6}).

Consequently, within this approximation, for the total number of photons emitted by a single electron and registered by a detector with the aperture $\theta_{\mathrm{D}}$ one finds
\begin{align}
  N_{ \sigma}(\theta_{0})  &= \frac{e_0^2}{4\hbar c} |\chi_g|^2 \left[\ln(D^2 + 1) - \frac{D^2}{D^2 + 1}\right]\nonumber
  \\
  &\mspace{45mu}\times\frac{(1 - e^{-L k_0\chi_0''\theta_{0} } )}{\chi_0''\theta_{0}},\label{eq:38}
\end{align}
where
\begin{align*}
  D = \frac{\theta_{\mathrm{D}}}{ \sqrt{\gamma^{-2} + \theta_{\mathrm{s}}^2 + |\chi_0'|}}.
\end{align*}

Eq.~(\ref{eq:38}) can be rewritten as
\begin{align}
  N_{\sigma}(\theta_{0}) = Q_{\text{PXR-EAD}}\xi, \label{eq:39}
  \\
  \xi = \frac{(1 - e^{-L k_0\chi_0''\theta_{0} } )}{\theta_{0}}. \label{eq:40}
\end{align}

As the last step we need to compare $Q_{\mathrm{PXR-EAD}}$ with an expression for $Q_{\mathrm{PXR}}$. For this Eq.~(1) of Ref.~\cite{Brenzinger1997} for the differential number of quanta $\partial^{2} N_{\mathrm{PXR}}/(\partial\theta_{x}\partial\theta_{y})$ emitted in the angular range $\theta_{x}$, $\theta_{y}$ can be employed. Consequently, we fix the same reflection $(400)$ and the same frequency of the emitted radiation $\omega_{\mathrm{B}}$, $\theta_{\mathrm{B}} = \pi/4$ as in our Eq.~(\ref{eq:38}). We also consider symmetrical transition geometry, when the planes $(100)$ are perpendicular to the crystal surface on which an electron beam is incident. In this situation $(\vec k'\cdot\vec N)/k_{0} = (\vec v_{0}\cdot\vec N)/v_{0} = \cos{\pi/4} = \sqrt{2}/2$. Moreover, we consider $L\gg L_{\mathrm{abs}}$. Then the total number of photons in the PXR peak is defined through the angular integration over $\theta_{x}$ and $\theta_{y}$ within the angular aperture of a detector $\theta_{\mathrm{D}}$. By performing this angular integration in Eq.~(1) of Ref.~\cite{Brenzinger1997} one can obtain the expression for the total number of quanta, which coincides with the value $Q_{\mathrm{PXR-EAD}}$ introduced in Eq.~(\ref{eq:39}). Consequently, the ratio $N_{\sigma}(\theta_{0})/N_{\mathrm{PXR}}$ is given with the parameter $\xi$ of Eq.~(\ref{eq:40}).

Finally, we will employ the parameters taken from the X-ray database \cite{StepanovXrayWebServer} for both numerical and analytical evaluation of the PXR radiation generated in a Si crystal by the reflection $(400)$
\begin{equation}
  \begin{aligned}
    &\hbar\omega_{\mathrm{B}} = 6.45\,\mathrm{keV},& &k_{0} = 3.29\times10^{8}\,\mathrm{cm}^{-1},
    \\
    &E = 900\,\mathrm{MeV},& &\theta_{\mathrm{s}}^{2} = 5.7\times10^{-5},
    \\
    &\chi_{0}' = -0.24\times10^{-4},& &\chi_{0}'' = 0.83\times 10^{-6},
    \\
    &\chi_{g}' = 0.12\times 10^{-4},& &\chi_{g}'' = 0.71\times 10^{-6},
    \\
    &\theta_{\mathrm{D}} = 10^{-2}, & & L = 1\, \mathrm{cm},
    \\
    &L_{\mathrm{abs}} = 3.7 \times 10^{-3}\, \mathrm{cm}, & & a_{0} = L_{\mathrm{abs}} / 2.
  \end{aligned}\label{eq:41}
\end{equation}

Lastly, we assume that two different electron beams with similar emmitances, but different transversal sizes and angular spreads are employed in the experiment, viz. $\epsilon = 5\times 10^{-6}$ $\mathrm{cm}\times \mathrm{rad}$, with $\Delta \theta_{\mathrm{e}} = 10^{-2}$ rad, $\Delta a = 5\times 10^{-4}$ cm and $\epsilon = 3\times 10^{-6}$ $\mathrm{cm}\times \mathrm{rad}$, with $\Delta \theta_{\mathrm{e}} = 3\times10^{-3}$ rad, $\Delta a = 10^{-3}$ cm.

During an actual experiment we suggest to measure the intensity of PXR radiation as a function of the beam incident angle $\theta_{0}$, which varies from $\theta_{0} = 0$ (grazing geometry) to $\theta_{0} = 0.3$ (transition geometry). Consequently, we predict the exponential increase of the PXR-EAD intensity when $\theta_{0}\to 0$.

In Fig.~\ref{fig:4} the comparison of the numerical evaluation and an analytical approximation Eq.~(\ref{eq:38}) is presented for the case when the crystal length $L\gg L_{\mathrm{abs}} = (k_{0}\chi_{0}'')^{-1}$. As follows from Fig.~\ref{fig:4} the number of photons of PXR-EAD significantly increases when the angle $\theta_{0} < [Lk_{0}\chi_{0}'']^{-1} = L_{\mathrm{abs}}/L$.

The parameter $\xi = N_{\mathrm{PXR-EAD}}/N_{\mathrm{PXR}}$ is proportional to the ratio $L/L_{\mathrm{abs}}$. Consequently, the increase of the radiation can be larger for lower photon frequencies, for which $L_{\mathrm{abs}}$ decreases. For example, in the considered geometry the reflection $(220)$ corresponds to $\hbar\omega_{\mathrm{B}} = 4.51$ keV and $L_{\mathrm{abs}} = 1.4\times 10^{-3}$ cm.

In Fig.~\ref{fig:5} we plot the dependence of the number of emitted quanta on the crystal length $L$, when the electron energy is fixed at $E = 900$ MeV. When the crystal length increases to $L_{\mathrm{opt}}\approx 2$ cm the PXR intensity reaches its maximum, while for crystal lengths $L > L_{\mathrm{opt}}$ the multiple electron scattering decreases the intensity in two ways. On the one hand, the multiple electron scattering decreases $D$ in Eq.~(\ref{eq:38}), which takes place also for the conventional transition geometry of PXR. On the other hand, it withdraws the electrons whose angle $\theta_{\mathrm{e}} \sim \theta_{\mathrm{s}} > |Lk_{0}\chi_{0}''|^{-1}$ from the layer $L_{\mathrm{abs}}$ and consequently the parameter $\xi$ in Eq.~(\ref{eq:39}) reduces. According to Eq.~(\ref{eq:8}) the angle $\theta_{\mathrm{s}}\sim 1/E$. Therefore, the optimal crystal length grows with the increasing electron energy as $L_{\mathrm{opt}}\sim 1/\theta_{\mathrm{s}} \sim E$, which additionally can increase the PXR-EAD intensity.

\section{Conclusions}
\label{sec:conclusions}

In the present work we investigated the influence of the geometry of an experiment on the PXR intensity. When the geometry of an experiment is changed from the transition to the grazing one, a significant increase of the PXR intensity takes place. The physical reason for this is that the electron beam propagates in a crystal in a layer parallel to the crystal-vacuum interface and the thickness of this layer is smaller than the absorption length of the X-ray radiation. Consequently, the emitted photons are not absorbed and the whole electron trajectory contributes to the formation of PXR radiation. In addition, we have demonstrated that the multiple-electron scattering does not prevent one to reach the predicted increase of the intensity.

\section{Acknowledgements}
\label{sec:acknowledgements}

ODS is grateful to C. H. Keitel, S. M. Cavaletto, S. Bragin and A. Angioi for helpful discussions.




\bibliographystyle{apsrev4-1}
\bibliography{pxr_nim}



\end{document}